\documentclass[a4paper,11pt]{article}
\usepackage{graphicx,amssymb,bm,latexsym}
\newcommand{\bea}{\begin{eqnarray}}
\newcommand{\ena}{\end{eqnarray}}
\newcommand{\vs}[1]{\vspace{#1 mm}}
\newcommand{\hs}[1]{\hspace{#1 mm}}
\renewcommand{\a}{\alpha}
\renewcommand{\b}{\beta}

\newcommand{\G}{\Gamma}
\renewcommand{\d}{\delta}
\newcommand{\e}{\epsilon}
\newcommand{\s}{\sigma}
\renewcommand{\t}{\theta}

\newcommand{\la}{\lambda}
\newcommand{\pa}{\partial}

\newcommand{\nn}{\nonumber\\}
\newcommand{\p}[1]{(\ref{#1})}
\newcommand{\lan}{\langle}
\newcommand{\ran}{\rangle}

\newcommand{\tg}{\tilde g}

\begin{document}

\begin{titlepage}

\begin{flushright}
KU-TP 055 \\
\end{flushright}

\vs{10}
\begin{center}
{\Large\bf Unitarity versus Renormalizability of Higher Derivative Gravity in $3D$}
\vs{15}

{\large
Kenji Muneyuki
and
Nobuyoshi Ohta\footnote{e-mail address: ohtan@phys.kindai.ac.jp}} \\
\vs{10}
{\em Department of Physics, Kinki University,
Higashi-Osaka, Osaka 577-8502, Japan}

\vs{15}
{\bf Abstract}
\end{center}

It has been suggested that new massive gravity with higher order terms in the curvature
may be renormalizable and thus a candidate for renormalizable quantum gravity.
We show that three-dimensional gravity that contains quadratic scalar curvature
and Ricci tensor is renormalizable, but those theories with special relation
between their coefficients including new massive gravity are not.

\end{titlepage}
\newpage
\setcounter{page}{2}

\section{Introduction}

Constructing quantum theory of gravity is one of the most important problems in
theoretical physics.
It has been known for some time that gravity is renormalizable in four dimensions
if one includes higher derivative terms~\cite{Stelle}. However there is a price.
The unitarity of the theory, which is one the most important properties of
any physical theory, is not preserved. So the theory has not been taken very seriously.

Recently a very interesting proposal has been made that the addition of such higher
order terms to three-dimensional gravity can keep the theory unitary and possibly
renormalizable if the coefficients are chosen appropriately~\cite{BHT1}.
The usual Einstein gravity does not have any propagating mode in three dimensions,
but the addition of these terms introduces propagating massive graviton around flat
Minkowski and curved maximally symmetric spaces (anti-de Sitter and de Sitter spaces).
A similar theory of massive graviton with higher derivative Lorentz-Chern-Simons (LCS) term has
long been known as topologically massive theory~\cite{Top}, but the theory violates parity.
In contrast, the new theory is a parity preserving theory, and it is called new massive gravity.
Since then, various aspects of the theory have been investigated.
Linearized excitations in the field equations were studied in~\cite{LS}.
Unitarity is proven for Minkowski space in~\cite{Oda1,Deser,GST}, whereas
it is discussed in \cite{BHT2} for maximally symmetric spaces.
A complete classification of the unitary theory for the most general action with arbitrary
coefficients of all possible terms is given in \cite{Ohta}.
A partial result of the unitarity condition on the flat Minkowski space is known
for the usual sign of the Einstein theory~\cite{NR}.

Though these kinds of theories have their own significance, it is also known that
such higher order terms are present in the low-energy effective theories of
superstrings. In this circumstance, these terms are regarded just as perturbative
corrections to the lowest order terms, and are not considered to be modes to be
quantized together. However, it is more appropriate to consider these terms together
if one would like to understand the quantum theory of gravity.
Also, there is some ambiguity in such theories due to the field redefinition.
If the approach of requiring unitarity and renormalizability determines
the coefficients to certain extent, it may cast some light on the superstrings themselves.

With such higher order terms, one may expect that the theory is renormalizable.
In fact, it has been argued that the topological massive gravity may be renormalizable
if suitable regularization is given~\cite{DY}, but this theory breaks parity and
the standard gauge-invariant regularization such as dimensional regularization
cannot be used. More recently, it was claimed that the new massive gravity is
renormalizable~\cite{Oda2}. However, there seem to be flaws in this argument, and
we will discuss where it fails.
There are also several arguments against renormalizability~\cite{BHTE},
and this is still under debate. It is important to settle the issue of whether
the gravitational theory is unitary and renormalizable or not.
In this paper we critically examine the renormalizability of the three-dimensional
gravity with quadratic curvature terms and try to clarify the situation.

Following the general discussions of renormalizability of four-dimensional
higher derivative gravity using Becchi-Rouet-Stora-Tyutin (BRST) symmetry~\cite{Stelle},
we study the quantum property of the three-dimensional theory. We examine the theory
with arbitrary coefficients for scalar and Ricci curvature squared but without a LCS term
because we assume the gauge-invariant dimensional regularization in our analysis, which
may not be used for parity-violating theory. This analysis shows that the theory is
renormalizable for general coefficients of these terms. However, we find that there
are two important exceptions to this. The renormalizability fails precisely for the cases of
unitary theories around flat Minkowski space, the new massive gravity and $R^2$ theory.
Thus, unfortunately, the unitarity and renormalizability do not seem to be compatible.
We would like to emphasize that it is important to settle this question
in the context of searching for quantum gravity.

\section{Higher Derivative Gravity in 3D}

Let us start with the action
\bea
S &=&\frac{1}{\kappa^2}\int d^3x \sqrt{- g} \Big[ \s R + \a R^2 +\b R_{\mu\nu}^2 \Big] \nn
& \equiv& \frac{1}{\kappa^2} \int d^3 x {\cal L}_{GMG},
\label{action}
\ena
where $\kappa^2$ is the three-dimensional gravitational constant,
$\a, \b$ and $\s (=0, \pm 1)$ are constants.
Though we can have topological mass term given by the gravitational LCS
term, we do not consider it in this paper for simplicity.

\subsection{Propagator}
\label{prop}

We define the fluctuation around the Minkowski background by
\bea
\tilde g^{\mu\nu}\equiv \sqrt{-g}\, g^{\mu\nu}
= \eta^{\mu\nu} + \kappa h^{\mu\nu}.
\label{fluc}
\ena
Substituting \p{fluc} into our action~\p{action}, we find the quadratic
term is given by
\bea
{\cal L}_2 \hs{-2}&=&\hs{-2} \frac{1}{4} h^{\mu\nu} \Big[ P^{(2)} (\b\Box+\s)
+ P^{(0,s)} \{(8\a+3\b)\Box-\s \}+ 2 P^{(0,w)} \{(8\a+3\b)\Box-\s \} \nn
&& \hs{10}
+\sqrt{2}(P^{(0,sw)}+P^{(0,ws)}) \{(8\a+3\b)\Box-\s \}\Big]_{\mu\nu,\rho\s}\Box h^{\rho\s},
\ena
where we have defined the projection operators as~\cite{NR}
\bea
P^{(2)}_{\mu\nu,\rho\s} \hs{-2}&=&\hs{-2} \frac12(\t_{\mu\rho}\t_{\nu\s}
+\t_{\mu\s}\t_{\nu\rho} -\t_{\mu\nu}\t_{\rho\s}), \nn
P^{(1)}_{\mu\nu,\rho\s} \hs{-2}&=&\hs{-2} \frac12(\t_{\mu\rho}\omega_{\nu\s}
+\t_{\mu\s}\omega_{\nu\rho}+\t_{\nu\rho}\omega_{\mu\s} + \t_{\nu\s}\omega_{\mu\rho}), \nn
P^{(0,s)}_{\mu\nu,\rho\s} \hs{-2}&=&\hs{-2} \frac12 \t_{\mu\nu}\t_{\rho\s}, ~~
P^{(0,w)}_{\mu\nu,\rho\s} =  \omega_{\mu\nu}\omega_{\rho\s},\nn
P^{(0,sw)}_{\mu\nu,\rho\s} \hs{-2}&=&\hs{-2} \frac1{\sqrt{2}} \t_{\mu\nu}\omega_{\rho\s},~~
P^{(0,ws)}_{\mu\nu,\rho\s} = \frac1{\sqrt{2}} \omega_{\mu\nu}\t_{\rho\s},
\ena
with
\bea
\t_{\mu\nu} = \eta_{\mu\nu}-\frac{\pa_\mu \pa_\nu}{\Box}, \qquad
\omega_{\mu\nu} = \frac{\pa_\mu \pa_\nu}{\Box}.
\ena
$P^{(2)}, P^{(1)}, P^{(0,s)}$ and $P^{(0,w)}$ are the projection operators onto spin
2, 1 and 0 parts, and they satisfy the completeness relation
\bea
( P^{(2)}+P^{(1)}+ P^{(0,s)}+ P^{(0,w)})_{\mu\nu,\rho\s}
= \frac12 (\eta_{\mu\rho} \eta_{\nu\s}+ \eta_{\mu\s} \eta_{\nu\s}),
\ena
on the symmetric second-rank tensors.

The BRST transformation for the fields is found to be
\bea
\d_B g_{\mu\nu} &=& -\d \la [ g_{\rho\nu}\pa_\mu c^\rho + g_{\rho\mu}\pa_\nu c^\rho
 + \pa_\rho g_{\mu\nu} c^\rho] 
, \nn
\d_B c^\mu &=& -\d\la c^\rho \pa_\rho c^\mu, \nn
\d_B \bar c_\mu &=& i \d\la\, B_\mu,\nn
\d_B B_\mu &=& 0,
\label{brst}
\ena
which is nilpotent. Here $\d\la$ is an anticommuting parameter.
We use the same gauge fixing as \cite{Stelle} using $\tg^{\mu\nu}$,
whose BRST transformation is given by
\bea
\d_B \tg^{\mu\nu} = \d\la (
\tg^{\mu\rho}\pa_\rho c^\nu + \tg^{\nu\rho}\pa_\rho c^\mu
-\tg^{\mu\nu}\pa_\rho c^\rho - \pa_\rho\tg^{\mu\nu} c^\rho)
\equiv \d\la {\cal D}^{\mu\nu}{}_\rho c^\rho.
\ena
The gauge fixing term and Faddeev-Popov (FP) ghost terms are concisely written as
\bea
{\cal L}_{GF+FP} &=& i \d_B [\bar c_\mu (\pa_\nu h^{\mu\nu}-\frac{a}{2}B^\mu)]/\d\la \nn
&=& - B_\mu \pa_\nu h^{\mu\nu}
- i \bar c_\mu \pa_\nu {\cal D}^{\mu\nu}{}_\rho c^\rho +\frac{a}{2} B_\mu B^\mu,
\label{gfgh}
\ena
where $a$ is a gauge parameter and the indices are raised and lowered with
the flat metric.

The simplest way to read off the propagator for the graviton is to first eliminate
the auxiliary field $B^\mu$ and look at the quadratic part. We find that it is given by
\bea
{\cal L}_{2,t} \hs{-2}&=&\hs{-2} \frac{1}{4} h^{\mu\nu} \Big[ P^{(2)} (\b\Box+\s)
+\frac{\Box}{a}P^{(1)} + P^{(0,s)} \{(8\a+3\b)\Box-\s \}
+ 2 P^{(0,w)} \{(8\a+3\b)\Box-\s +\frac{1}{a} \} \nn
&& \hs{10}
+\sqrt{2}(P^{(0,sw)}+P^{(0,ws)}) \{(8\a+3\b)\Box-\s \}\Big]_{\mu\nu,\rho\s}\Box h^{\rho\s},
\label{quad}
\ena
Using the completeness property and the orthogonality of the projection operators,
we find the propagator is given by
\bea
D_{\mu\nu,\rho\s}(k) \hs{-2}&=&\hs{-2} \frac{1}{(2\pi)^3} \Big[ \frac{P^{(2)}}{k^2(\b k^2-\s)}
+ \frac{P^{(0,s)}}{ k^2 \{(8\a+3\b)k^2+ \s\}} \nn
&& -\frac{a}{2 k^2} \{2P^{(1)} +2 P^{(0,s)} + P^{(0,w)} -\sqrt{2}(P^{(0,sw)}+P^{(0,ws)}) \}
\Big]_{\mu\nu,\rho\s}.
\label{gpropagator}
\ena
We shall take the Landau gauge
given by $a=0$. Since the theory is invariant under the general coordinate transformation,
this does not cause any problem, but simplifies the discussions considerably~\cite{Stelle}.
In this gauge, we have
\bea
\pa_\mu h^{\mu\nu}=0.
\label{landau}
\ena
Note that the above propagator satisfies $k^\mu D_{\mu\nu,\rho\s}(k)=0$ in this gauge.
Also, the propagator damps as $k^{-4}$ for large momentum.
As a result, as we argue later, the theory becomes renormalizable. However, there
are two important exceptional cases: $8\a+3\b=0$ and $\b=0$.
The former case corresponds to the new massive gravity
recently discussed extensively~\cite{BHT1}. Although this theory is found to be
unitary either for $8\a+3\b=0$ or $\b=0$~\cite{Deser}-\cite{Ohta},
the following power counting argument fails precisely in these cases
and the theory is not renormalizable.
For $\s=0$, in addition to $8\a+3\b=0$ discussed in \cite{Deser}, the propagator
cannot be obtained in the usual sense, and the following argument for the
renormalizability does not apply.

\subsection{Slavnov-Taylor identity}
\label{stid}

If we introduce the Grassmann-odd source $K_{\mu\nu}$ and the Grassmann-even source $L_\mu$,
we have the BRST-invariant action
\bea
I_{sym}[h_{\mu\nu}, \bar c_\a, c^\b, K_{\mu\nu}, L_\rho]
\hs{-2} &=& \hs{-2} \int d^3 x[ {\cal L}_{GMG}+{\cal L}_{GF+FP}
+ K_{\mu\nu}{\cal D}^{\mu\nu}{}_\rho c^\rho - L_\mu c^\nu \pa_\nu c^\mu] \nn
& \equiv &\hs{-2}  \int d^3 x \; {\cal L}_{sym}.
\ena
The BRST invariance follows from \p{brst}, \p{gfgh} and the nilpotency of
the BRST transformation.

The generating functional of Green's functions is given by
\bea
Z[J_{\mu\nu}, \bar \eta_\a, \eta^\b, K_{\mu\nu}, L_\rho]
\hs{-2} &=& \hs{-2} \int [dh][d\bar c][dc] \exp\left(i\int d^3 x [ {\cal L}_{sym}
+J_{\mu\nu} h^{\mu\nu} + \bar\eta_\a c^\a + \bar c_\a \eta^\a] \right) \nn
\hs{-2} &\equiv& \hs{-2} \exp\left( i W[J_{\mu\nu}, \bar \eta_\a, \eta^\b, K_{\mu\nu}, L_\rho]
 \right),
\label{generating}
\ena
where $J_{\mu\nu}$, and $\bar\eta_\a$ and $\eta^\a$ are Grassmann-even and Grassmann-odd
sources, respectively.
The BRST invariance of the functional~\p{generating}
\bea
0 = \int [dh][d\bar c][dc] \d_B \exp\left(i\int d^3 x [ {\cal L}_{sym}
+J_{\mu\nu} h^{\mu\nu} + \bar\eta_\a c^\a + \bar c_\a \eta^\a] \right),
\ena
implies that
\bea
\left\lan \int d^3 x \left[ J_{\mu\nu} {\cal D}^{\mu\nu}{}_\rho c^\rho
+ \bar \eta_\mu c^\nu \pa_\nu c^\mu +i\frac{1}{a} \eta_\mu \pa_\nu h^{\mu\nu}
 \right] \right\ran = 0,
\ena
where the field $B_\mu$ is eliminated by its field equation.
This yields the Slavnov-Taylor identity
\bea
\int d^3 x \left[ J_{\mu\nu} \frac{\d W}{\d K_{\mu\nu}} -\bar\eta_\mu \frac{\d W}{\d L_\mu}
+\frac{i}{a} \eta^\mu \pa_\nu \frac{\d W}{\d J_{\mu\nu}} \right]=0.
\ena
The equations of motion for the FP ghost is
\bea
\pa_\nu \frac{\d W}{\d K_{\mu\nu}} + i \eta_\mu=0.
\ena
As usual, the effective action is defined by
\bea
\tilde \G[h^{\mu\nu}, \bar c_\a, c^\b, K_{\mu\nu}, L_\rho]
\equiv W[J_{\mu\nu}, \bar \eta_\a, \eta^\b, K_{\mu\nu}, L_\rho]
-\int d^3 x \left[J_{\mu\nu} h^{\mu\nu} +\bar\eta_\a c^\a+\bar c_\a \eta^\a \right].
\label{effective}
\ena
It follows from \p{generating} that
\bea
h^{\mu\nu} = \frac{\d W}{\d J_{\mu\nu}},~~~
c^\mu = \frac{\d W}{\d \bar\eta_\mu}, ~~
\bar c_\mu = - \frac{\d W}{\d \eta^\mu}.
\ena
The relations dual to these are
\bea
J_{\mu\nu} = - \frac{\d \tilde\G}{\d h^{\mu\nu}}, ~~~
\bar \eta_\a = \frac{\d \tilde\G}{\d c^\a}, ~~
\eta^\a = -\frac{\d \tilde\G}{\d \bar c_\a}.
\ena
We further define
\bea
\G=\tilde\G+ \int d^3 x \frac{1}{2a} (\pa_\nu h^{\mu\nu})^2.
\ena
With the help of the relations
\bea
\frac{\d \G}{\d K_{\mu\nu}}= \frac{\d W}{\d K_{\mu\nu}},~~~
\frac{\d \G}{\d L_\mu}= \frac{\d W}{\d L_\mu},
\ena
and the ghost field equation
\bea
\pa^\nu \frac{\d \G}{\d K_{\mu\nu}} - i \frac{\d \G}{\d \bar c^\mu} = 0,
\ena
the Slavnov-Taylor identity reduces to
\bea
\label{stid1}
\int d^3 x \left[ \frac{\d \G}{\d h^{\mu\nu}} \frac{\d \G}{\d K_{\mu\nu}}
+ \frac{\d \G}{\d c^\mu} \frac{\d \G}{\d L_\mu}\right]=0,
\ena

The $n$-loop part of the effective action is denoted by $\G^{(n)}$.
The effective action is a sum of these terms:
\bea
\G = \sum_{n=0}^\infty \G^{(n)}.
\ena
Suppose that we have successfully renormalized the effective action up to
$(n-1)$-loop order. Write
\bea
\G^{(n)} = \G^{(n)}_{\rm finite} + \G^{(n)}_{\rm div}.
\ena
If we insert this breakup into Eq.~\p{stid1} and keep only the terms which are
of $n$-loop order, we get
\bea
\int d^3 x \left[ \frac{\d \G^{(n)}_{\rm div}}{\d h^{\mu\nu}}\frac{\d \G^{(0)}}{\d K_{\mu\nu}}
+ \frac{\d \G^{(n)}_{\rm div}}{\d c^\mu}\frac{\d \G^{(0)}}{\d L_\mu}
+ \frac{\d \G^{(0)}}{\d h^{\mu\nu}}\frac{\d \G^{(n)}_{\rm div}}{\d K_{\mu\nu}}
+ \frac{\d \G^{(0)}}{\d c^\mu}\frac{\d \G^{(n)}_{\rm div}}{\d L_\mu} \right] \nn
= - \int d^3 x  \sum_{i=0}^n \left[
\frac{\d \G^{(n-i)}_{\rm finite}}{\d h^{\mu\nu}}\frac{\d \G^{(i)}_{\rm finite}}{\d K_{\mu\nu}}
+ \frac{\d \G^{(n-i)}_{\rm finite}}{\d c^\rho}\frac{\d \G^{(i)}_{\rm finite}}{\d L_\rho}
\right].
\label{stid2}
\ena
Since each term on the right-hand side of \p{stid2} remains finite as $\e(=d-3)\to 0$
in the dimensional regularization, while each term on the left-hand side contains
a factor with a pole in $\e$, each side of the equation must vanish separately.
This leads to
\bea
\int d^3 x \left[ \frac{\d \G^{(0)}}{\d K_{\mu\nu}} \frac{\d }{\d h^{\mu\nu}}
+ \frac{\d \G^{(0)}}{\d L_\la}\frac{\d }{\d c^\la}
+ \frac{\d \G^{(0)}}{\d h^{\mu\nu}}\frac{\d}{\d K_{\mu\nu}}
+ \frac{\d \G^{(0)}}{\d c^\la}\frac{\d}{\d L_\la} \right] \G^{(n)}_{\rm div}
=0.
\label{stid3}
\ena
This identity will be used in later discussions of renormalizability.

\subsection{Renormalizability}

Under the expansion~\p{fluc}, the Einstein term gives graviton vertices with
two derivatives, and curvature square terms give those with four derivatives.
Consider arbitrary Feynman diagrams. We use the following notations.

$V_2$: the number of graviton vertices with two derivatives from the $R$ term.

$V_4$: the number of graviton vertices with four derivatives from the $R^2$ term.

$V_c$: the number of ghost-antighost-graviton vertices with two derivatives.

$V_K$: the number of $K$-graviton-ghost vertices.

$V_L$: the number of $L$-ghost-ghost vertices.

$I_h$: the number of internal-graviton propagators.

$I_c$: the number of internal-ghost propagators.

$E_h$: the number of external gravitons.

$E_c$: the number of external ghosts.
\\
Since the graviton propagator behaves as $k^{-4}$ and the FP ghost propagator as $k^{-2}$,
we are led by the standard power counting to the degree of divergence of
an arbitrary diagram:
\bea
D=3L-4 I_h -2 I_c+4 V_4+2 V_2+2 V_c+V_K+V_L.
\ena
Using the relation
\bea
L=I_h+I_c-(V_4+V_2+V_c+V_K+V_L-1),
\ena
we get
\bea
D=3-I_h+I_c+V_4-V_2-V_c-2V_K-2V_L.
\ena
We further use the topological relation
\bea
2V_c+V_K+2V_L=2I_c+E_c+E_{\bar c},
\ena
to obtain
\bea
D=3-(I_h-V_4)-V_2-\frac32 V_K- V_L- \frac12 (E_c+ E_{\bar c}).
\label{pc}
\ena
Now the ghost vertex contained in the FP ghost term in \p{gfgh}, upon partial integration,
can be rewritten as
\bea
i[\pa_\rho \pa_\mu \bar c_\nu \cdot c^\nu h^{\mu\rho}
+\pa_\mu \bar c_\nu \cdot c^\nu \pa_\rho h^{\mu\rho}
+\pa_\mu \bar c_\nu \cdot c^\mu \pa_\rho h^{\nu\rho}].
\ena
In the Landau gauge in which we have \p{landau}, the last two terms do not couple to the
propagator. Also integration by parts in the remaining term can be used to move
the derivative onto the ghost using the gauge condition:
\bea
i \pa_\rho \pa_\mu \bar c_\nu \cdot c^\nu h^{\mu\rho}
\approx i \bar c_\nu \pa_\rho \pa_\mu c^\nu h^{\mu\rho}.
\ena
As a result, in one-particle irreducible (1PI) diagrams, each external ghost and antighost
carries two factors of external momentum~\cite{Stelle}.
The resulting degree of divergence of an arbitrary 1PI diagram is then
\bea
D^{(1PI)} = 3-(I_h-V_4)-V_2-\frac32 V_K- V_L- \frac52 (E_c + E_{\bar c}).
\ena
We note that $I_h-V_4 \geq 0$ for 1PI diagrams. Consequently we find that the possible
divergences are restricted; those with external ghosts and antighosts have
$D^{\rm (1PI)}\leq -2$, those with the external $K$ and ghost
$D^{\rm (1PI)} \leq -1$, and those with $L$ and two ghosts have $D^{\rm (1PI)} \leq -3$.
Hence, we have
\bea
\frac{\d \G^{(n)}_{\rm div}}{\d c^\la}
= \frac{\d \G^{(n)}_{\rm div}}{\d K^{\mu\nu}}
= \frac{\d \G^{(n)}_{\rm div}}{\d L_\la}=0.
\label{nodiv}
\ena
The Slavnov-Taylor identity~\p{stid3} then reduces to
\bea
\int d^3 x \frac{\d \G^{(0)}}{\d K^{\mu\nu}} \frac{\d \G^{(n)}_{\rm div}}{\d \tg^{\mu\nu}}
=0.
\ena
Together with \p{nodiv}, this implies that $\G^{(n)}_{\rm div}$ is gauge invariant.
Consequently $\G^{(n)}_{\rm div}$ are local gauge-invariant functionals of $\tg^{\mu\nu}$
with zero and two derivatives (up to three).
This allows only the counterterms of the Einstein and the cosmological terms.
Hence the theory is super-renormalizable.

However we note that there are very important exceptional cases to these arguments:
$8\a+3\b=0$ and $\b=0$. As noted in Sec.~\ref{stid}, the behavior of the propagator
becomes $k^{-2}$ even in the Landau gauge and the degree of divergence is changed to
\bea
D^{(1PI)} = 3+I_h+V_4-V_2-\frac32 V_K- V_L- \frac52 (E_c + E_{\bar c}).
\ena
Diagrams are more divergent as the numbers of the vertices and propagators of the graviton
are increased, and the theory is not renormalizable.

\section{Discussions and conclusions}

In this paper we have studied whether higher derivative gravity in three dimensions
is renormalizable or not. As it turns out, the general theory is renormalizable,
but if we restrict the coefficients to special values $8\a+3\b=0$ or $\b=0$,
then the convergence property of the graviton propagator becomes worse, and the power
counting indicates that the theory is not renormalizable.
Unfortunately, these are precisely the cases when the theories are unitary at
the tree level; the former case is known as new massive gravity, and the latter
is a special case of $f(R)$ gravity.

In new massive gravity, it is known that the condition $8\a+3\b=0$ eliminates
the propagating spin 0 mode. From the graviton propagator~\p{gpropagator}, we see
that it is the spin 0 component of the graviton that gives this bad behavior to
the propagator. If the scalar mode decouples from all physical quantities,
then this would not spoil the renormalizability of the theory. However, this is not the case.
Even if the scalar mode is excluded in the initial states, this mode enters
physical processes through the interactions so that it produces divergences which
cannot be renormalized.
Only if there is some kind of local gauge invariance to ensure the decoupling of
the scalar mode can the theory be renormalizable.
However, there is no such invariance in our present higher derivative gravity
in three dimensions and the theory is not renormalizable with the above restriction
on the coefficients.

There has been an argument that claims the new massive gravity is renormalizable~\cite{Oda2}.
We find that there are two points that need careful consideration.
The first is that the author introduces mass terms which break the gauge invariance
of the theory and derives the graviton propagator, which has bad behavior
and makes the resulting power counting invalid.
However, it was argued that these parts decouple and one can ignore these bad behaviors;
only the $k^{-4}$ behavior of the spin 2 component is taken.
This does not seem to be justified because there is no invariance that ensures the decoupling
of the scalar mode in this theory when interactions are taken into account.
Moreover the gauge invariance or the invariance under the general coordinate transformation
is explicitly broken by the mass term, and it is expected that there arises an inconsistency
in the spin-2 part with nonlinear interactions.
Second, the author claims that the massless limit gives the new massive gravity so
that the latter itself is renormalizable. However the massless limit is subtle
in general field theories, and we expect that this procedure is dubious.
If such manipulation were allowed, we could have proved the renormalizability
of the Yang-Mills theory including the mass term.
Considering these points, we believe that this ``proof'' cannot be justified and
conclude that the new massive gravity is not renormalizable though the general theory is.
Thus unfortunately the new massive gravity does not give the quantum thoery of gravity,
but our analysis opens the possibility that if there is some symmetry which can eliminate
the scalar mode, there is a possibility of finding renormalizable and unitary theory.

\section*{Acknowledgement}

We would like to thank S. Deser, A. Ishibashi and P. K. Townsend
for very useful discussions.
This work was supported in part by the Grant-in-Aid for
Scientific Research Fund of the JSPS (C) No. 20540283, No.
21$\cdot\,$09225 and (A) No. 22244030.

\end{document}